\documentclass[
    ,final            
  ]
  {aipproc}
\usepackage{graphicx, amsmath, amssymb, epsf, bm,fdag}
\layoutstyle{6x9}

\begin{document}

\title{A $k_T$-dependent sea-quark density for the 
 \textsc{Cascade}  Monte Carlo  event generator }


\author{F.~Hautmann}{
  address={Department of Theoretical Physics, University of Oxford, Oxford OX1 3NP}
}
\author{M.~Hentschinski}{
  address={Instituto de F\'isica Te\'orica UAM/CSIC, Universidad Aut\'onoma de Madrid, E-28049 Madrid},
altaddress={Physics Department, Brookhaven National Laboratory, Upton, NY 11973, USA}
 }

\author{H.~Jung}{
  address={Deutsches Elektronen Synchrotron, D-22603 Hamburg, Germany}
  ,altaddress={CERN, Physics Department, CH-1211 Geneva 23} 
}

\begin{abstract} 
Parton-shower  event generators that go beyond the collinear-ordering  approximation 
at small $x$ have  so far  included    
only gluon and valence quark channels at  transverse momentum dependent  level.    
We describe results of recent work to include effects of the  sea-quark
  distribution with explicit dependence on the transverse
  quark-momentum.  This sea-quark density is then applied to the 
  description of forward Z -production. The $qq^* \to Z$ matrix
  element (with one off-shell quark) is calculated in an explicit
  gauge invariant way, making use of high energy factorization. The
  $k_T$-factorized result has been implemented into the CCFM Monte-Carlo
  \textsc{Cascade} and a numerical comparison with the 
  $qg^*\to Zq$ matrix element has been carried out.
\end{abstract}

\maketitle


\section{Introduction}
Transverse momentum dependent parton distribution arise at small $x$
naturally as a consequence of high energy factorization and
BFKL-evolution \cite{bfkl}. A formulation of high energy factorization
which is in accordance  with conventional collinear factorization is
provided by $k_T$-factorization as defined in
\cite{ktfac,Catani:1994sq}: at high center of mass energies, the
resummation of high energy logarithms (BFKL) can be brought into a
form consistent with conventional collinear resummation (DGLAP). The
CCFM evolution equation \cite{ccfm}, which interpolates between DGLAP
and BFKL evolution, allows then for a Monte Carlo
implementation of $k_T$-factorization, which is realized by the Monte Carlo
event generator \textsc{Cascade} \cite{cascade}.

Based on the principle of color coherence, the CCFM parton shower
describes only the emission of gluons, while real quarks emissions are
left aside. Note that such an approximation is in principle justified
as    enhanced regions of phase space for  $x\to 1$ and $x\to
0$  are dominated by  gluonic dynamics at leading logarithmic 
level. For the description of
unintegrated parton densities this implies that the CCFM evolution
describes only the distinct evolution of  unintegrated gluon and 
valence quarks \cite{forwdjet} while non-diagonal transitions between
quarks and gluons are absent. 
On the other hand, it is necessary to
include quark emissions into the parton shower to take into 
account  subleading  effects.  Also the determination
of $k_T$-dependent hard matrix elements is affected by the absence of
quark emissions:  gluon-to-quark splittings  needed for the
generation of seaquarks  are to  be included into  matrix elements,  which
complicates their  structure, see {\it e.g.}  \cite{DYmichal}.  In order 
 to supplement the CCFM evolution with quarks we present
in this contribution a definition of a partonic matrix element
involving off-shell quarks and the definition of an off-shell
sea-quark density. For the latter we restrict here  to the
case where the gluon-to-quark splitting occurs as the last evolution
step. To test our procedure  numerically, we apply it   to the 
 production of a $Z$-boson in the forward direction at LHC
energies. For a detailed discussion  we refer to \cite{HHJ,HHJp}.

\section{Definition of a $q_T$-dependent sea-quark}
Let us take the approach of generalizing $k_T$-factorization to the case of
quarks  by  mimicking  the already existing gluonic case.  There off-shell
gauge invariance is ensured through a reformulation of QCD at high
center of mass energies in terms of effective degrees of freedom,
reggeized gluons. The latter coincide in their on-shell limit with
conventional collinear QCD gluons, while one uses for the general
off-shell case effective vertices which contain additional induced
terms  ensuring off-shell gauge invariance\footnote{For an
  effective action approach to the reggeized gluon formalism see
  \cite{reggeized,Hentschinski:2011tz}.}. In complete analogy one can construct a
reggeized quark formalism for the description of the high energy limit
of scattering amplitudes with quark exchange in the $t$-channel
\cite{reggequarks}. As (reggeized) quark exchanges are 
 suppressed by powers of $s$  in comparison
to (reggeized) gluon exchanges, they
generally do not occur in the high energy resummation of total
cross-sections. They can however be used as a starting point for the
construction of an off-shell factorization of matrix elements which
are limited to quark exchange in the $t$-channel.  This is
the case for the $g^*q \to Zq$ matrix element for which we 
construct \cite{HHJ} an off-shell factorization into an off-shell matrix element
$q^*q \to Z$ and a corresponding gluon-to-quark splitting
function.  
Making use of the reggeized quark formalism,  
we  arrive  at the off-shell  partonic cross-section $qq^*\to Z$,
\begin{align}
  \label{eq:sigmahat}
  \hat{\sigma}(x_1x_2 s, M_Z^2, {\bf q}^2 ) &= 
\sqrt{2} G_F M_Z^2 (V_q^2 + A_q^2)  \frac{\pi}{N_c} \delta(x_1x_2 s  - {\bf q}^2 - M_Z^2).
\end{align}
Here $x_1 \sqrt{s}$ and $x_2 \sqrt{s}$ are the light-cone momenta of
the two incoming quarks, while ${\bf q}$ denotes the transverse
momentum of the off-shell quark; $M_Z$ denotes the mass of the
$Z$-boson, while $\sqrt{2} G_F M_Z^2 (V_q^2 + A_q^2)$ describes the
coupling of the $Z$-boson to the quark.  
Eq.~\eqref{eq:sigmahat} agrees for ${{\bf q}}^2 \to 0$ with the $qq\to
Z$ cross-section, while  the splitting function is 
obtained as a pure color factor  if the reggeized quark formalism is
taken literally --- a  crude approximation which can be traced back to
non-conservation of energy at the level of high-energy leading logarithms.  
  It is  possible however \cite{HHJ} to supply energy conservation 
while maintaining all desired gauge invariance properties. Doing so we
arrive at the  $k_T$-dependent \cite{Catani:1994sq}  gluon-to-quark splitting function
\begin{align}
  \label{eq:ktsplitt_def}
  {P}_{qg} \bigg(z, \frac{{\bf k}^2 }{ {\bf q}^2 }\bigg) = 
T_R \left( 
            \frac{ {\bf q}^2 }{\,  {\bf q}^2  + z(1-z){\bf k}^2}
\right)^2 
\left[
        {(1-z)^2 + z^2 } + 4z^2 (1-z)^2 \frac{  {\bf k}^2 }{{\bf q}^2}
\right], 
\end{align}
where ${\bf k}$ and ${\bf q}$ denote
transverse momenta of the off-shell gluon and quark respectively, while
$z$ denotes the fraction of gluon light cone momentum which is carried
on by the $t$-channel quark.
 In the on-shell limit ${\bf k}^2 
\to 0$ eq.~\eqref{eq:ktsplitt_def} reduces to the 
collinear  DGLAP splitting function $T_R [z^2 + (1-z)^2]$ at lowest order;  in the 
high-energy limit however it factorizes correctly the finite ${\bf k}$ dependence to 
all orders.  
  At the next-to-leading logarithmic accuracy $\alpha_s ( \alpha_s \ln x )^n $,  
  the  $q_T$-dependent sea-quark density is   defined by
\begin{align}
  \label{eq:seaqark}
  \mathcal{Q}^{\text{sea}} (x, {\bf q}^2, \mu^2) &=   \frac{1}{{\bf q}^2}
   \int\limits_x^1  \frac{d z}{z}
\int \limits_0^{{\bf k}_{\text{max}}^2}
 d {\bf k}^2 
  \frac{\alpha_s(\mu^2)}{2 \pi} {P}_{qg} \left(z, \frac{ {\bf k}^2}  {{\bf q}^2}\right) 
\mathcal{G}\left(\frac{x}{z}, {\bf k}^2, \bar{\mu}^2   \right), 
\end{align}
with ${\bf k}^2_{\text{max}} = {  \mu^2}/z - { {\bf q}^2 }/{(z (1-z))} $.
For the  scale $\bar{\mu}$  of the unintegrated gluon density, 
we  will  investigate 
 two possible choices. 
 One choice, closer in spirit to 
 inclusive calculations  
\cite{Catani:1994sq}, is  to set simply $\bar{\mu}^2 = \mu^2$.  A 
choice based on angular ordering, which is more natural from the point
of view of the CCFM evolution, is given by
$ \bar{\mu}^2 = {\bf q}^2/(1-z)^2 +  {\bf k}^2/(1-z)$.
The complete $q_T$-factorized cross-section for forward $Z$-production
 is given by 
\begin{align}
  \label{eq:heXsecbetter_fertig}
   \sigma_{pp \to Z}^{q_T-\text{fact.}}(s, M_Z^2) &=  \sum_f
 \int_{0}^1 {d x_1}  \int _0^\infty {d {\bm q}^2 } \int_{0}^1 {d x_2}  \,  \hat{\sigma}^f_{qq^* \to Z}\,  \mathcal{Q}^{\text{sea}}(x_1, {\bf q}^2, \mu^2){Q}^f(x_2, \mu^2)
\end{align}
with ${Q}^f(x_2, \mu^2)$ the  valence quark distribution,  implemented as in the first paper 
of~\cite{forwdjet}.    

\section{Numerical test of the off-shell factorization}
\label{sec:numerics}
\begin{figure}[h]
  \centering

  \parbox{.5\textwidth}{\includegraphics[width=.5\textwidth]{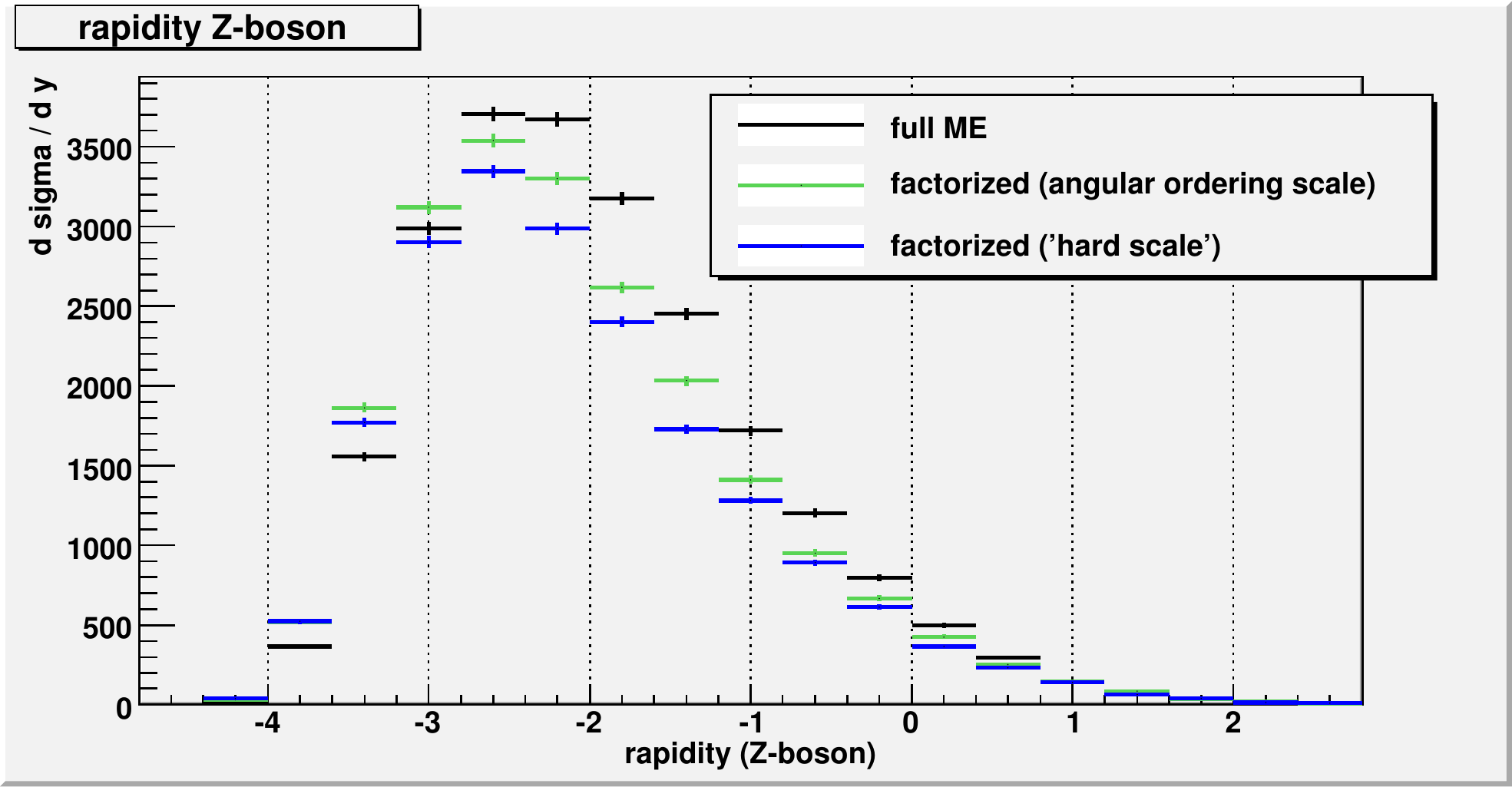}}
\parbox{.5\textwidth}{ \includegraphics[width = .5 \textwidth]{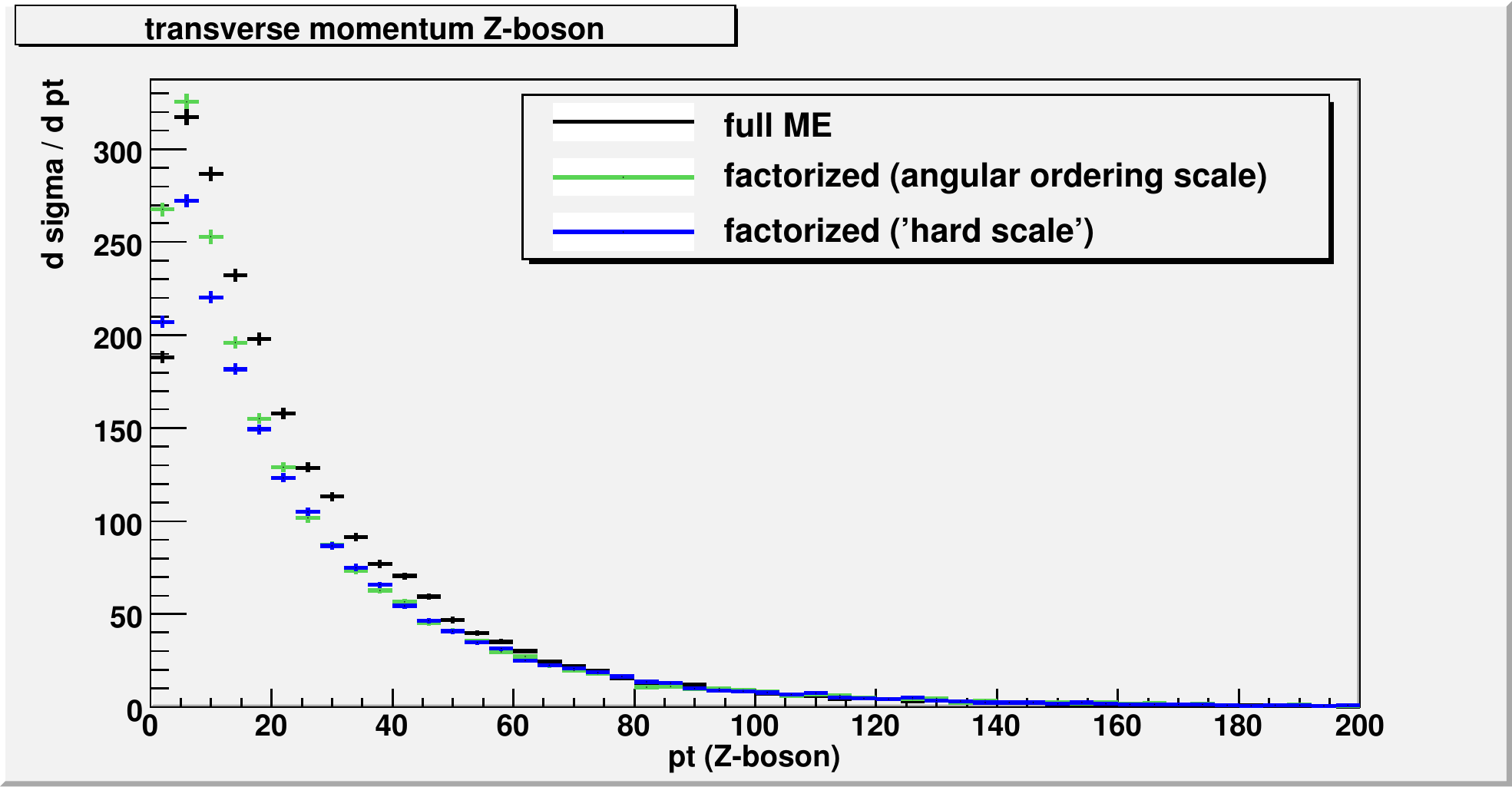}}
  
  \caption{\small}
  \label{fig:ptandrap1}
\end{figure}
\begin{figure}[h]
  \centering

\parbox{.5\textwidth}{ \includegraphics[width=.5\textwidth]{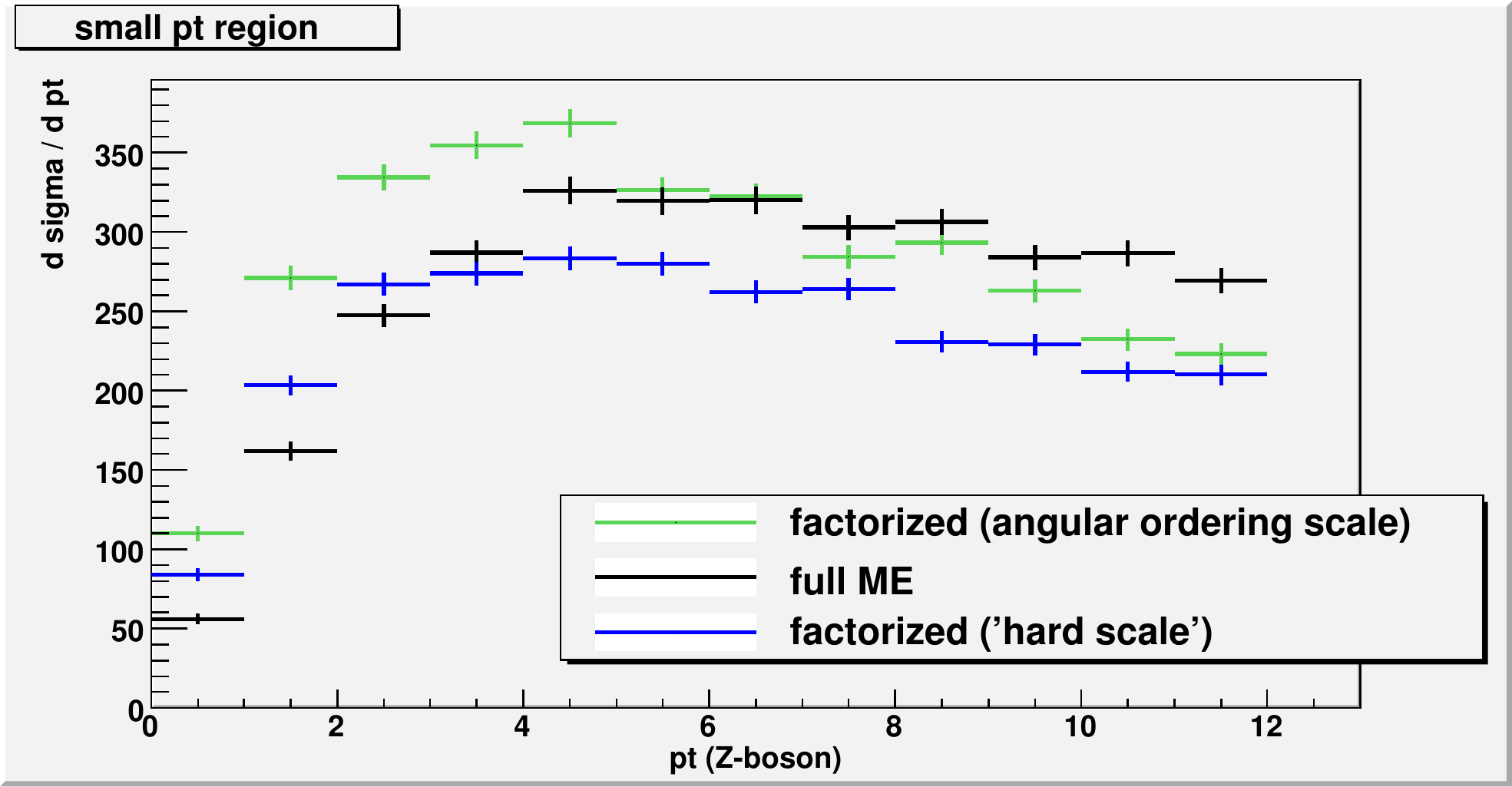}}
\parbox{.5\textwidth}{ \includegraphics[width = .5 \textwidth]{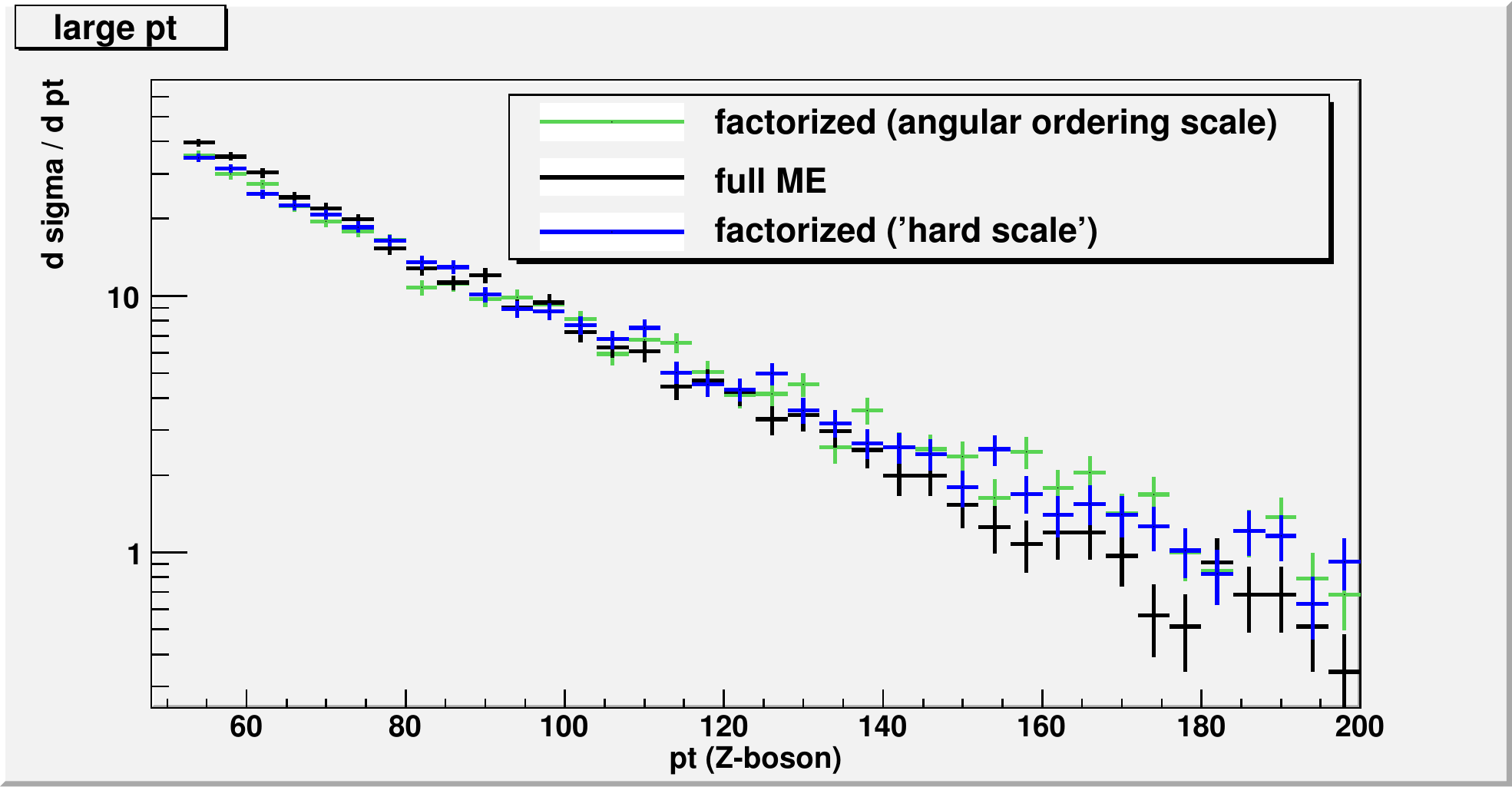}}
  
  \caption{\small}
  \label{fig:ptandrap2}
\end{figure}

 We  next  compare  numerically  the factorized expression
eq.~\eqref{eq:heXsecbetter_fertig} with the result of 
implementing the $g^*q \to Zq$
off-shell cross-section    \cite{HHJ,Marzani:2008uh}
in the \textsc{Cascade}  Monte Carlo  
generator \cite{cascade}. The results are summarized 
in fig.~\ref{fig:ptandrap1} and fig.~\ref{fig:ptandrap2}. 
For the factorized result both choices for the hard scale
$\bar{\mu}^2$ have been implemented. The  differences
between the matrix element (ME) and  the factorized  results are    
discussed in  \cite{HHJp}.  We note that 
  the factorized
result  contains the $t$-channel contribution, while  the 
$s$-channel contribution  is sub-leading both in
the collinear and high energy limit.  The factorized expression also contains a
kinematical approximation~\cite{HHJ,reggequarks} in comparison to the full $g^*q \to Zq$
cross-section, which leads to a small, but finite correction.

In summary, we achieved an off-shell factorization of the $g^*q \to
Zq$ process, where the factorized expression can  both analytically
and numerically be shown to coincide with the full $g^*q \to Zq$
cross-section up to a finite remainder, sub-leading in the collinear
and the high energy limit.


\begin{theacknowledgments}
  M.H. acknowledges support from the German Academic Exchange Service
  (DAAD), DESY, the U.S. Department of Energy under contract number
  DE-AC02-98CH10886 and a BNL ``Laboratory Directed Research and
  Development'' grant (LDRD 12-034).
\end{theacknowledgments}



\bibliographystyle{aipproc}   


\IfFileExists{\jobname.bbl}{}
 {\typeout{}
  \typeout{******************************************}
  \typeout{** Please run "bibtex \jobname" to optain}
  \typeout{** the bibliography and then re-run LaTeX}
  \typeout{** twice to fix the references!}
  \typeout{******************************************}
  \typeout{}
 }

\end{document}